\documentclass{article}

\usepackage{arxiv}

\usepackage[utf8]{inputenc} 
\usepackage[T1]{fontenc}    
\usepackage{hyperref}       
\usepackage{url}            
\usepackage{booktabs}       
\usepackage{amsfonts}       
\usepackage{nicefrac}       
\usepackage{microtype}      
\usepackage{lipsum}         
\usepackage{graphicx}
\usepackage{doi}
\usepackage{comment}
\usepackage{csquotes}
\usepackage{hyperref}
\usepackage{subcaption}
\newcommand{\ie}{{\em i.e.}}

\newcommand{\etal}{{\em et al.}}
\usepackage{amsmath}
\usepackage{cleveref} 

\usepackage{algorithmic}
\usepackage{textcomp}
\usepackage{xcolor}
\usepackage{balance}

\title{TrojanForge: Generating Adversarial Hardware Trojan Examples Using Reinforcement Learning}

\date{}

\author{ \href{https://orcid.org/my-orcid?orcid=0000-0002-0134-8418}{\includegraphics[scale=0.06]{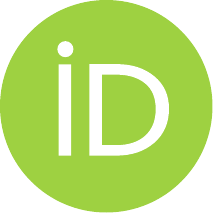}\hspace{1mm}Amin ~Sarihi} \\
	Klipsch School of Electrical\\ and Computer Engineering\\
	New Mexico State University\\
	\texttt{sarihi@nmsu.edu} \\
	\And
	\href{https://orcid.org/0000-0002-3741-0201}{\includegraphics[scale=0.06]{orcid.pdf}\hspace{1mm}Peter ~Jamieson} \\
	Department of Electrical\\ and Computer Engineering\\
	Miami University\\
	\texttt{jamiespa@miamioh.edu}\\
        \And
	\href{https://orcid.org/0000-0003-2647-2797}{\includegraphics[scale=0.06]{orcid.pdf}\hspace{1mm}Ahmad ~Patooghy} \\
	Department of Computer Systems Technology\\
	North Carolina A\&T  State University\\
	\texttt{apatooghy@ncat.edu}\\
	\And
	\href{https://orcid.org/0000-0001-8027-1449}{\includegraphics[scale=0.06]{orcid.pdf}\hspace{1mm}Abdel-Hameed A. ~Badawy} \\
	Klipsch School of Electrical\\ and Computer Engineering\\
	New Mexico State University\\
	\texttt{badawy@nmsu.edu} \\}

\renewcommand{\headeright}{}
\renewcommand{\undertitle}{}
\renewcommand{\shorttitle}{TrojanForge: Adversarial Hardware Trojan Examples Using Reinforcement Learning}

\begin{document}
\maketitle

\begin{abstract}
	The Hardware Trojan (HT) problem can be thought of as a continuous game between attackers and defenders, each striving to outsmart the other by leveraging any available means for an advantage. Machine Learning (ML) has recently played a key role in advancing HT research. Various novel techniques, such as Reinforcement Learning (RL) and Graph Neural Networks (GNNs), have shown HT insertion and detection capabilities. HT insertion with ML techniques, specifically, has seen a spike in research activity due to the shortcomings of conventional HT benchmarks and the inherent human design bias that occurs when we create them. This work continues this innovation by presenting a tool called \textquote{TrojanForge}, capable of generating HT adversarial examples that defeat HT detectors; demonstrating the capabilities of GAN-like adversarial tools for automatic HT insertion. We introduce an RL environment where the RL insertion agent interacts with HT detectors in an insertion-detection loop where the agent collects rewards based on its success in bypassing HT detectors. Our results show that this process helps inserted HTs evade various HT detectors, achieving high attack success percentages. This tool provides insight into why HT insertion fails in some instances and how we can leverage this knowledge in defense.
\end{abstract}


The economics of the silicon supply chain has driven silicon vendors to adopt a globalized fabless business model to curb costs. Most design layouts are sent overseas to be manufactured at semiconductor foundries~\cite{bhunia2018hardware}. This has introduced a significant increase in revenue for many companies, where the share of the fabless semiconductor firms’ revenue has reached 33\% of the entire chip industry in 2020 from just 7.6\% in 2000~\cite{xing2023global}. Nevertheless, this success comes with a security cost imposed to such chips i.e., outsourced designs can be subject to unwanted malicious design modifications such as Hardware Trojans (HT), \ie, malicious embeddings that can be introduced at the chip design or manufacturing stage. An HT has a triggering component that defines the conditions under which the HT is activated, and upon activation, the payload of the HT, which is the malicious action or behavior, can cause a variety of threats that compromise the Confidentiality, Integrity, or Availability of hardware devices, such as Denial of Service (DoS), information leakage, reducing transistor age, and thermal leakage~\cite{xue2020ten}. 

To study the impact of (HTs), researchers have primarily used the dataset available at \url{Trust-hub.org}~\cite{shakya2017benchmarking}. HT benchmarks are essential assets because they provide opportunities to study the impact of various HTs and the development of HT detection methods. Despite the valuable effort to create these benchmarks for the research community, several associated shortcomings exist. Krieg~\cite{krieg2023reflections} has studied these benchmarks from correctness, maliciousness, stealthiness, and persistence standpoints. He deduced that only three out of 83 studied HT circuits are usable in real-world scenarios. The major issues were pre-synthesis and post-synthesis discrepancies, unsatisfiable trigger conditions, and incorrect original designs. In addition to this, the Trusthub benchmark set is restricted in size and diversity, hindering the ability to train efficient HT detectors that rely on large training datasets. Last but not least, the inherent human bias in HT insertion in Trusthub circuits is another serious problem that undermines the benchmark's quality. These shortcomings have urged the community to find alternative solutions to these benchmarks in the past few years by developing various automated HT insertion tools~\cite{cruz2018automated,gohil2022attrition,sarihi2022hardware,gohil2024attackgnn}.

\begin{figure*}[ht]
  \centering
  \includegraphics[width=\linewidth]{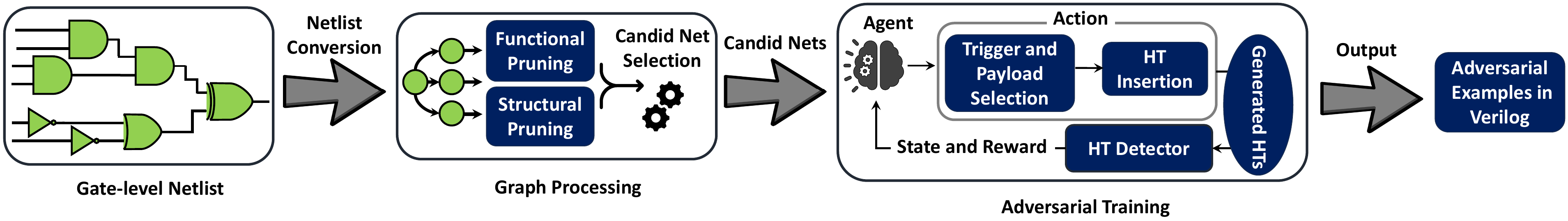}
  \caption{Insertion Flow of TrojanForge.}
  \label{fig:flow}  
\end{figure*}

In the Machine Learning (ML) domain, adversarial examples are crafted inputs that are manipulated to mislead machine learning models and force them to return wrong predictions. 
Following the same paradigm, \textbf{we seek inspiration from Generative Adversarial Networks (GANs) to generate novel and more challenging HT insertions. Our proposed Reinforcement Learning (RL) agent, resembling the GAN’s generator component, generates HT instances intending to evade HT detectors.} We propose the~\textquote{TrojanForge}
framework in which an HT inserter tool operates in conjunction with HT detectors (the GAN’s Discriminator) in a loop and tries to collect the maximum positive reward by generating HT instances that are not detectable. The generated HT instances can be found in~\cite{githubGitHubAminsarihiTrojanForgeAdversarialHardwareTrojanExamples}. Figure~\ref{fig:flow} shows an overview of our proposed framework where a selection of trigger candidate nets are picked from a netlist and are passed to an RL HT insertion agent. (More details will be explained in Section \ref{sec:methodology}.)

This work’s threat model assumes HT insertion either at design or manufacturing time, \ie, a malicious employee who has access to the state-of-the-art reverse-engineering techniques in a foundry, reverse-engineers a design to obtain the netlist, and embeds HTs into this netlist. Similar threat assumptions have been made~\cite{gohil2022attrition,gohil2022deterrent}. We assume that the adversary uses existing HT detectors or can develop them. 


This work makes the following contributions:
\begin{itemize}
    \item We develop an RL-based HT generator and employ various HT detector(s) in a GAN-like loop to generate adversarial HT examples.  
    \item We use functional and structural pruning techniques to narrow the HT trigger selection process.
    \item We utilize the Jaccard Similarity Index~\cite{pandit2011comparative} to diversify candidate nets for HT insertion. This helps the HT inserting agent to proceed when HTs are hard to insert. The metric also provides insight into designing and defending HT insertion.
    \item We demonstrate how the choice of payload can impact the stealthiness of HTs and lead to a false sense of security from the defense perspective.  
\end{itemize}

The rest of this paper is organized as follows. Section~\ref{sec:background} reviews background work in HT insertion and detection. Section~\ref{sec:methodology} delves into our proposed framework and its details. Section~\ref{sec:results} analyzes the results of this new tool. Finally, Section~\ref{sec:conclusion} concludes the paper.

\section{Review of HT Insertion \& Detection Literature}
\label{sec:background}


Initial attempts to gather an HT benchmark were realized in TrustHub, where a list of 96 benchmarks are available to use~\cite{shakya2017benchmarking,salmani2013design}. Despite the valuable novelty, the dataset suffers from shortcomings such as limited size and diversity~\cite{krieg2023reflections}. 
Various studies have been conducted to address Trusthub’s shortcomings. Cruz~\etal~\cite{cruz2018automated} offers an automated HT generation tool that inserts HTs based on entered parameters from the user. They include the total number of trigger nets, the number of rare trigger nets, the number of inserted HT instances, and the type of payload. The location of a payload is selected randomly. 

Sarihi~\etal~\cite{sarihi2022hardware} introduced an RL-based tool in which an agent explores circuits and takes five different positional actions to insert HT instances in a way that maximizes the sum of collected rewards. The agent receives rewards according to the number of engaged circuit inputs in the HT activation test vector. Gohil~\etal~\cite{gohil2022attrition} proposed another HT insertion RL tool, ATTRITION. ATTRITION sets a signal probability threshold for trigger candidates, and the RL agent is rewarded proportional to the size of \textquote{compatible} trigger rare nets it finds. Compatible rare nets are a set of rare nets that can be activated with a single test vector. Nozawa~\etal~\cite{nozawa2021generating} degrade the performance of ML-based HT detectors by introducing adversarial HT examples. The approach changes the structure of the netlist while preserving its functionality. These changes tamper with the feature values of the circuit and, subsequently, cause a circuit misclassification by the detector. The authors use Trusthub HTs to showcase the success of their method. Sarihi~\etal~\cite{sarihi2024seeker} uses ABC, an open-source synthesis tool, to restructure graphs while preserving their functionality. The goal is to introduce a new benchmark with unseen data points. The authors demonstrate that current state-of-the-art HT detectors cannot detect most of the proposed benchmarks. Gohil~\etal~\cite{gohil2024attackgnn} introduced AttackGNN, a tool that integrates ABC with an RL agent to generate adversarial examples that mislead GNNs (graph neural networks) which are used in four hardware security applications. The RL agent uses a combination of graph restructuring actions to achieve the highest attack rates. 

Saravanan~\cite{saravanan2023revisiting}~\etal~propose a rare net selection technique that has two main parts. First, the authors use the structural information of circuits such as the number of unique primary inputs and gates that form the input logic cone of a rare net to analyze each rare net. Second, random sampling with a tree-based search is used to uncover overlapped trigger combinations and score them based on their quality.

As our proposed framework needs a discriminator (HT detector) tool, in the rest of this section, we review HT detection literature. 
TGRL~\cite{pan2021automated} is an RL framework that generates test vectors by tweaking the input bit patterns to activate a set of rare nets. The algorithm shows higher trigger coverage compared to MERO; however, neither the test vectors are released, nor the vectors are tested on real HT benchmarks. DETERRENT~\cite{gohil2022deterrent} also employs RL to find the largest set of rare nets that are compatible~\cite{gohil2022deterrent}. The authors discuss how to prune some actions for a faster convergence; however, they do not consider rare net pruning. Sarihi~\etal~\cite{10406091} introduces three detection proposals based on SCOAP (Sandia Controllability and Observability Analysis Program parameters) and signal switching activity to generate test vectors for ISCAS-85 benchmarks. The authors combine all the test vectors~\cite{sarihi2024trojan} and demonstrate their efficacy against HT implementations. Hasegawa~\etal~\cite{hasegawa2022r} introduces a robust ML HT detector called R-HTDetector trained with 51 features extracted from TrustHub benchmarks. The detector aims to increase resiliency against the adversarial gate modification attacks proposed in~\cite{nozawa2021generating}.
\section{TrojanForge}
\label{sec:methodology}

In this section, we describe our proposed framework, TrojanForge, consisting of the tools that conduct a GAN-like HT insertion. The first step in the process is to obtain the gate-level netlist of the design. Next, a Python script converts the netlist to an equivalent acyclic-directed graph representation. A graph consists of a set of vertices and edges $G = (V, E)$, where $V$ is a set of vertices (gates), and $E$ is a set of edges (nets), such that $E \subseteq ( V \times V$). Before the RL agent starts processing the graph for HT insertion, a set of candidate nets will be used to build the HT triggers. We will use the terms nets and edges interchangeably in this work. We also assume that our HTs are combinational circuits with an XOR payload similar to~\cite{cruz2018automated,gohil2022attrition,sarihi2022hardware}. 
\subsection{Rare Net Pruning}

Since each design consists of thousands of nets, choosing candidate nets for HT trigger circuitry is costly. In this section, we propose a heuristic to reduce the search space. This process starts with a set of 50,000 random test vectors simulated on each netlist, and we record the resulting internal net values. A filter removes candidate nets with a switching probability greater than $th_{sw}$, similar to~\cite{cruz2018automated,gohil2022attrition}. The remaining nets will be referred to as \textit{rare nets} denoted as set $T$. This initial pruning step is essentially a probability-based filtering. Next, we execute two sequential steps: \textit{Functional Pruning} described in Section~\ref{subsec:H} and \textit{Structural Pruning} in Section~\ref{subsec:V}. Both steps aim to further narrow down the candidate set, $T$, reducing the search space for the RL agent and shortening its training time.

\subsubsection{\textbf{Functional Pruning}}
\label{subsec:H}

\begin{figure}[!t]
  \centering
  \includegraphics[width=.65\linewidth]{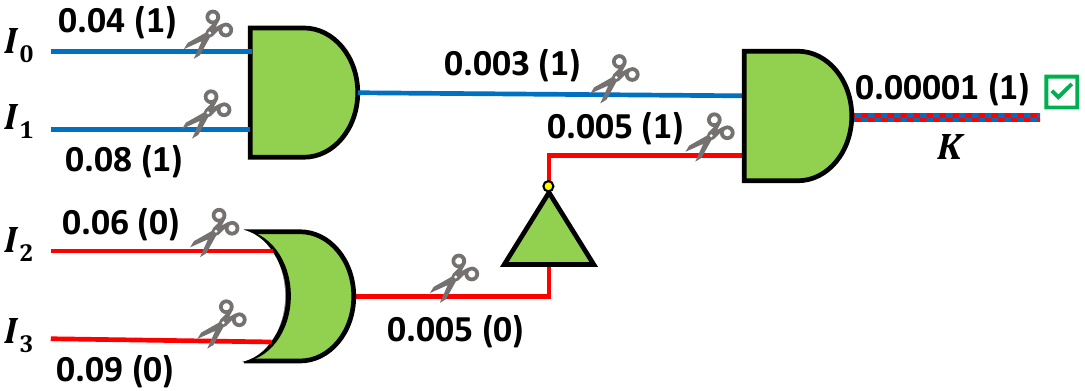}
  \caption{Applying Functional Pruning two rare paths. The final candidate net $K$ is selected to represent the rare nets. }
  \label{fig:Fprune}  
\end{figure}

This first technique is called \textit{Functional Pruning}, and it removes redundant rate nets based on their functional dependency. Figure~\ref{fig:Fprune} shows an example of such a process for two logical paths colored in blue and red. The rare nets follow the format of $ SW (RV)$ where $SW$ is the switching probability, and $RV$ is the rare value. Equation~\ref{eq:func} explains the circuit’s functionality
\begin{equation}
    K = (I_0~AND~I_1)~AND~(NOT~(I_2~OR~I_3))
    \label{eq:func}
\end{equation}
where $I_i$ are the primary inputs and $K$ is the output net. Value $I = 1100 $ guarantees that all nets of both paths are assigned with their rare values, making them trigger candidates. However, these nets drive the output $K$, \ie, the output net becomes $1$ if and only if all rare nets of blue and red paths get their rare values. Thus, we keep $K$ and remove all other rare nets due to functional dependency. We denote the obtained set as $F_{pruned}$ where $F_{pruned} \subseteq T$. Functional Pruning starts from scanning the graph from the higher logic levels towards the lower logic levels to find a net, similar to $K$, with a switching probability of lower than $th_{sw}$. For each $K$, we find the nets with functional dependencies in its input logic cone. Lastly, we try to activate $K$ with a test vector derived from the rare paths and prune the nets. The nets that were discarded in the process will not be revisited.

\subsubsection{\textbf{Structural Pruning}}
\label{subsec:V}

The next technique called \textit{Structural Pruning}, is motivated by when rare nets within the same logic level have similar structural features and functionality. 

\begin{table}[!t]
    \centering
    \caption{A snapshot of net properties in $c3540$}
    \begin{tabular}{|c|c|c|c|c|c|c|}
        \hline
        \textbf{Edge} & \textbf{RV} & \textbf{SW} & \textbf{Level} & \textbf{CC0} & \textbf{CC1}&\textbf{HTS} \\ \hline \hline
        (436, 1431) & 0 & 0.03 & 5 & 11 & 3 & 0.727  \\ \hline
        (440, 1439) & 0 & 0.03 & 5 & 11 & 3 & 0.727 \\ \hline
        (468, 1491) & 0 & 0.03 & 5 & 11 & 3 & 0.727 \\ \hline
        (472, 1499) & 0 & 0.03 & 5 & 11 & 3 & 0.727 \\ \hline
    \end{tabular}
    \label{tab:vprune}
\end{table}

Before starting this process, we compute the HT trigger susceptibility parameter~\cite{sebt2018circuit} for the already functionally pruned nets. Equation~\ref{HTS1} describes this parameter:
\begin{equation}
    HTS(Net_i)=\frac{|CC1(Net_i)-CC0(Net_i)|}{Max(CC1(Net_i),CC0(Net_i))}
    \label{HTS1}
\end{equation}
where $HTS$ is the HT trigger susceptibility parameter of $Net_i$. $CC0(Net_i)$ and $CC1(Net_i)$ are the combinational controllability $0$ and $1$ of $Net_i$, respectively. $HTS$ stems from the fact that low-switching nets have mainly a high difference between their controllability values. $HTS$ falls within $[0,1)$, and higher values correlate with lower activity on the net. The Structural Pruning process involves deriving a feature vector for each trigger candidate in the graph. Table~\ref{tab:vprune} displays a snapshot of such properties for four nets in $c3540$ after a functional pruning where Edge denotes the edge id, $RV$ shows the rare value, $SW$ is the switching probability, and $Level$ is the logic level of the net. Also, for each edge, we derive its functionality from its logic input cone. As can be seen from the table, these nets display similar structural behaviors resembling the behavior of a bus in the circuit. While they might not share the exact input logic cone, their functionality shows a high resemblance. Hence, instead of including these nets in the candidate rare net set, we only pick one as a representative net in this pruning process. This step leaves us with a set of nets with unique $HTS$ values upon completion. We denote this set as $S_{pruned}$ where $S_{pruned} \subseteq F_{pruned}$.

Finally, we pick a random set of trigger nets from $S_{pruned}$ to be named $M$ where $M\subseteq S_{pruned}$. As we plan to insert 5-input HTs, all possible subsets of size $5$ out of $\lvert M \lvert$ are equal to $\binom{\lvert M \lvert}{5}$. The complexity of the problem grows with bigger  $\lvert M\lvert $. Nevertheless, the attacker can cause a significant threat with even a smaller $M$. The value of $M$ is selected to include ample insertion opportunities while keeping the trigger selection problem computationally reasonable.
\subsection{Adversarial Training}
\label{sec:adv}
When configured for 5-input HT insertion, the RL agent uses a vector \( A \), defined as \( A = [a_0, a_1, a_2, a_3, a_4] \), to represent the action space, where each \( a_i \in [0, \lvert M \rvert - 1] \). A state space vector $S$ is the sorted version of the action space, \ie, $S = sort([a_0,a_1,a_2,a_3,a_4])$. If a repeated action (trigger) is selected, it will be replaced by a random one. Next, the payload target net $P$ is selected from all the nets except the ones in $M$. $P$ has a higher logic level than all of the trigger candidates.
The payload selection constraints ensure no combinational loops in the final circuit. We use an ATPG algorithm (used in~\cite{sarihi2022hardware}) to ensure that inserted HTs are functional. 

To complete the GAN-like feedback step, the infected graph is tested against a detector to evaluate the quality of the HT insertion. The HT detector could be either an ML-based detector that runs graph structural features for detection purposes or a test-based HT detector that checks circuit outputs for any deviation from the golden ones. Subsequently, the detector outputs a binary value informing the agent whether it can detect the generated HT instance. From the perspective of the inserting agent, we consider three rewarding scenarios in this situation:
\begin{enumerate}
    \item If the RL agent fails to insert a functional HT, the agent is rewarded a negative number $R_{negative}$. This negative reinforcement forces the agent to avoid such actions.
    \item When an inserted HT is functional and detected, the agent is rewarded `0'. This helps the agent distinguish between successful and unsuccessful HT insertions.
    \item When an inserted HT is functional and bypasses the detection, the RL is rewarded with a large positive number $R_{positive}$.
\end{enumerate}
In our experiments, $R_{positive}$ is substantially larger than $R_{negative}$ since actions that lead to such a reward are rarer and should be reinforced in the RL. $R_{positive} $’s value varies from one circuit to another since each circuit is treated as a new RL environment with its characteristics and challenges.

One possible solution for TrojanForge is that each generated HT instance could be tested against thousands of test vectors. In such a case, the RL agent’s 1000 HT insertion training attempts can lead to over a few million detection checks, making the training flow slow and impractical. To overcome this issue, we feed the detection test vectors to the circuit before the start of the HT generation process and record the number of compatible \textit{quin-triggers}, \ie, a set of five unique nets that can be activated together. We refer to the number of activated quin-triggers as the $Q-coverage$. After each HT insertion, the RL detector checks whether the current quin-trigger exists in its Q-coverage and rewards the agent accordingly. Searching for quin-triggers introduces a complexity of $O(N)$ where $N = \binom{\lvert M \lvert}{5}$. This provides a significant speedup over testing each insertion against thousands of test vectors.
\subsection{Special Case: Incompatible Triggers}

While the current flow leads to successful and stealthy HT insertions, there are some circuits in which the tool cannot insert HTs. The reason for this is having a high similarity in the input logic cones of the candidate nets (meaning they share either the same upstream logic or similar functional upstream logic). In such a case, the ATPG algorithm fails to find test vectors that trigger HT instances since shared inputs in the triggers’ logic cone impose conflicting values. 

To better understand the problem, we have created a heuristic metric called \textit{Jaccard Similarity Index} ($JSI$). This parameter informs the agent about the similarity of input logic cones of candidate trigger nets. The $JSI$ for two nets and their upstream logic cones is computed using Equation~\ref{eq:JSI}:
\begin{equation}
   JSI =  \frac{\lvert L_{i}\cap L_{j}\lvert }{\lvert L_{i}\cup L_{j}\lvert }
   \label{eq:JSI}
\end{equation}
where $L_i$ and $L_j$ contain the list of PIs (primary inputs) in the input logic cone of the $i_{th}$  and the $j_{th}$ triggers, respectively. The operator $\lvert x \lvert$ returns the number of items in the list $x$. $JSI$ ranges between $[0,1]$ where higher values indicate more upstream logic cone similarity between triggers, possibly leading to less viable insertions. When the set of trigger candidate nets has a high $JSI$ value, our heuristic approach is to replace some candidate nets with \textit{regular nets} to lower $JSI$. The set of regular nets is denoted as $R$ where $R \subseteq E - T$. When there are indications of high $JSI$, we set a threshold $JSI_{th}$ and group nets with similar JSI  thresholds together. We keep one net from each group with the smallest input logic cone size and replace the remaining nets with regular nets. The replaced nets do not share any input logic cones with the existing set. This process introduces diversity into the trigger choice. We note that this step of the proposed framework emerged from the analysis of our experiments. The $JSI$ metric and our heuristic approach here suggest that the metric might be used as a design goal for a circuit to help protect it from HT insertions.

\section{Experimental Results}
\label{sec:results}

In this section, our focus is to demonstrate the capabilities of our GAN-like adversarial HT insertion system and demonstrate how it can evade HT detection. For our experiments, we use an AMD EPYC 7702P 64-core CPU with 512GB of RAM to pre-process graphs, train, and test the RL agents. RL agents are trained with the Proximal Policy Optimization (PPO) algorithm~\cite{schulman2017proximal} from the Stable Baselines  library~\cite{raffin2019stable}. We target ISCAS-85~\cite{iscas85} benchmark circuits for our study and use NetworkX~\cite{hagberg2008exploring} to process graphs. The RL episode length is set to 25, and with each environment reset, we assign a new payload net to enable the RL agent to experiment with different payloads, thus preventing it from becoming stuck with ones that cannot effectively propagate the HTs’ impact to the circuit outputs. Publicly available test vectors from DETERRENT~\cite{gohil2022deterrent} and three sets of test vectors from Sarihi~\etal~\cite{10406091} are used for HT detection. In this study, $M=20$. If the attacker wants to explore higher $M$ values, RL algorithms can deal with large, complex environments. We will report our results based on the pairing of our HT insertion with the appropriate discriminator. We note that the nature of our circuit benchmarks is both small and simple and does not reflect modern circuit design. This is a noted shortcoming in both HT detection and insertion space, which we are helping to improve with ML-based tools such as the TrojanForge. 


\subsection{JSI and Trigger  Compatibility}
\label{subsec:JSI}

Table~\ref{tab:quintets} shows Q-coverage (explained in Section~\ref{sec:adv}) for the combinational ISCAS-85 benchmark circuits (listed in each row) when using random test vectors, $D1$, $D2$, $D3$ from~\cite{10406091}, and DETERRENT (labeled in each column, respectively).
\begin{table}[!t]
    \centering
    \caption{Q-coverage of HT detectors for ISCAS-85 circuits}
    \scalebox{0.95}{
        \begin{tabular}{|c|c|c|c|c|c|}
            \hline
            \textbf{Benchmark} & \textbf{Random} & \textbf{D1} & \textbf{D2} & \textbf{D3} & \textbf{DETERRENT} \\ \hline \hline
            $c880$ & 600 & 1909 & 1865 & 1149 & N/A\\ \hline
            $c1355$ & 0 & 0 & 0 & 0 & N/A\\ \hline
            $c1908$ & 7 & 14 & 14 & 14  & N/A \\ \hline
            $c2670$ & 30 & 173 & 163 & 147 & 1 \\ \hline
            $c3540$ & 45 & 26 & 70 & 80 & N/A \\ \hline
            $c5315$ & 45 & 222 & 364 & 190 & 5646 \\ \hline
            $c6288$ & 55 & 5989 & 6098 & 631 & 7020\\ \hline
            $c7552$ & 25 & 388 & 501 & 6 & 536\\ \hline
        \end{tabular}
    }
    \label{tab:quintets}
\end{table}
DETERRENT only includes test vectors for four circuits ($c2670, c5315, c6288, c7552$), and it provides higher Q-coverage than other detectors except for the case of $c2670$. While Gohil~\etal~\cite{gohil2022deterrent} shows 100\% four-input coverage with 8 test vectors for $c2670$, this is not the case with our candidate selection strategy. Such interpretations could lead to false security interpretations and vulnerability exploitation by attackers. By selecting 20 candidate trigger nets ($\lvert M \lvert=20$), up to 15,504 checks are needed for each insertion by the detector; however, circuit structures and detectors’ imperfections lead to significantly fewer checks.

\begin{figure}[!t]
  \centering

  \includegraphics[scale=.65]{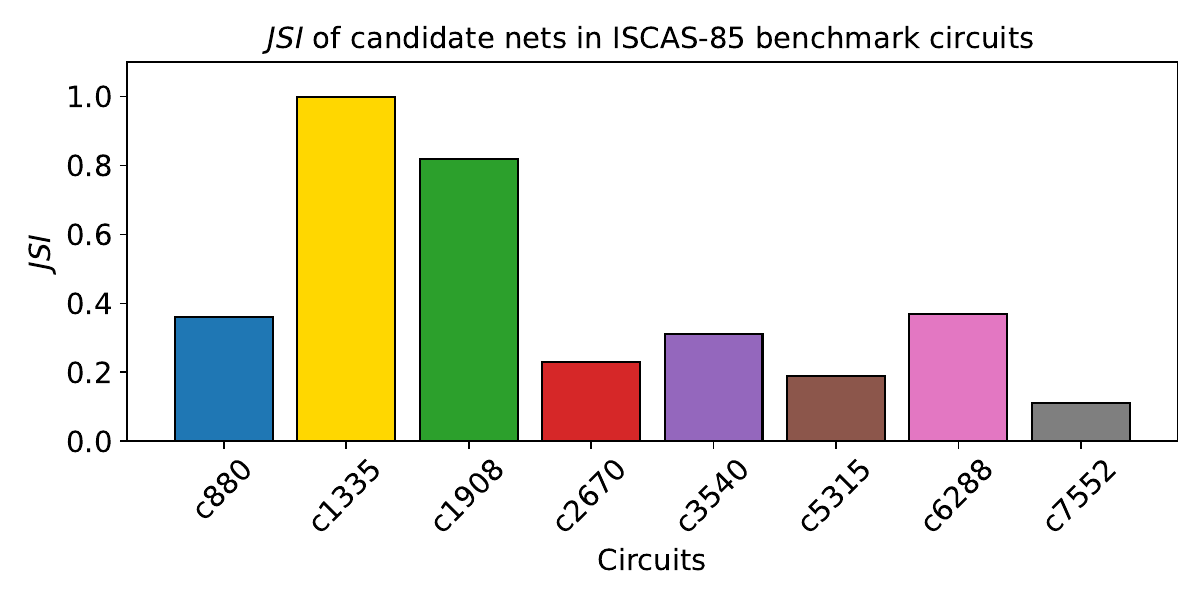}
  \caption{$JSI$ of set $T$  for each ISCAS-85 circuit.}
  \label{fig:JSI}  
\end{figure}

Two circuits in Table~\ref{tab:quintets} show unusually low Q-coverage, namely $c1355$ and $c1908$. The former is a 32-bit single-error-correcting circuit, and $c1908$ is a 16-bit error detector/corrector. Upon examining the Verilog code of $c1355$, a recurring pattern emerges that involves XOR operations on inputs, followed by NAND operations and assignments to the outputs. This pattern leads to highly similar logic cones for the candidate nets. To further investigate this, we compute set $T$ and its $JSI$ for all eight circuits in Figure~\ref{fig:JSI}. As can be seen, these two circuits demonstrate remarkably high $JSI$ values. $JSI$ values accompanied by the Q-coverage table suggest that stealthy HT insertion can be challenging for these two circuits. 
\subsection{HT Insertion in TrojanForge}

\begin{figure}[!t]
  \centering
  \includegraphics[scale=.6]{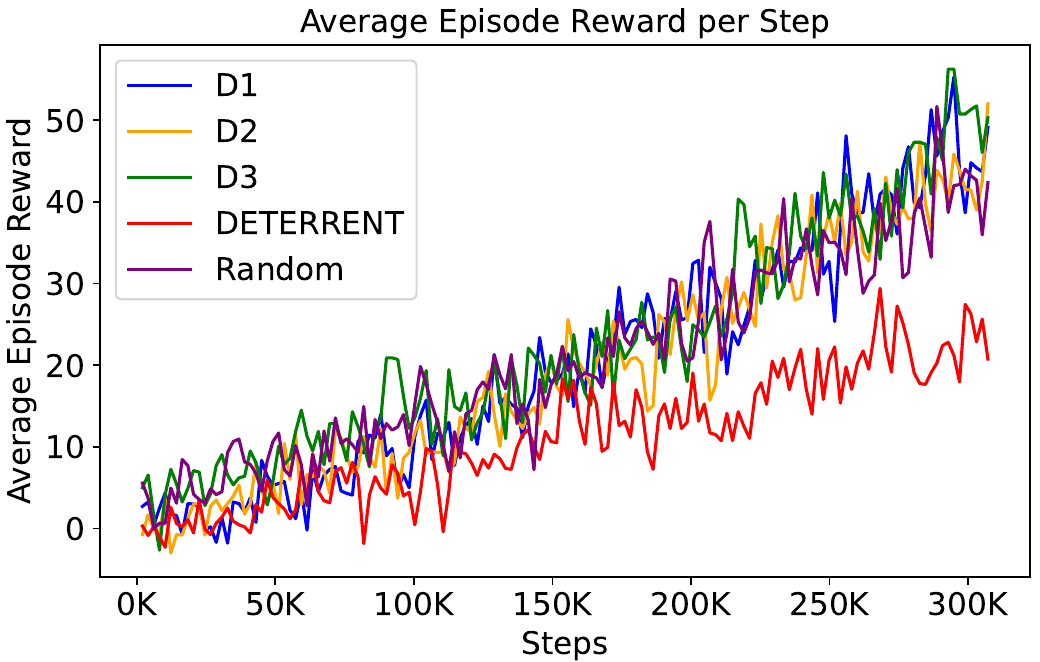}
  \caption{Average Episode Reward per Step of the RL agent when inserting against D1, D2, D3, Deterrent, and Random detectors for $c7552$.}
  \label{fig:reward}  
\end{figure}

We train the RL agent for each circuit with five different training strategies, \ie, each insertion strategy is specialized to beat one of the five HT detectors used. The training is done for 300K timesteps for each circuit, \ie, 300K insertion attempts. During the training process, we ensure that the RL agent learns and accumulates positive rewards. Figure \ref{fig:reward} shows the average episode reward per step of the RL agent when inserting against D1, D2, D3, DETERRENT, and Random detectors for a sample circuit $c7552$. The scale of accumulated rewards is directly proportional to the Q-coverage in Table \ref{tab:quintets}. The reward accumulation trends in other circuits, except in cases where the agent was not able to defeat the detector(explained later in the session), are very similar to Figure~\ref{fig:reward} which demonstrates robust learning. Although the reward trajectory shows an upward trend in all insertion scenarios, we stop training at 300K timesteps since we can achieve significant attack success rates within this time window (later explained in this section). After training, we enter the generation phase and store the first 100 generated HTs. To evaluate the quality of these HTs, we test them against all detectors except for the strategy against which they were trained.

\begin{figure*}[t]
    \centering
    \begin{subfigure}{0.31\textwidth}
        \centering
        \includegraphics[width=\textwidth]{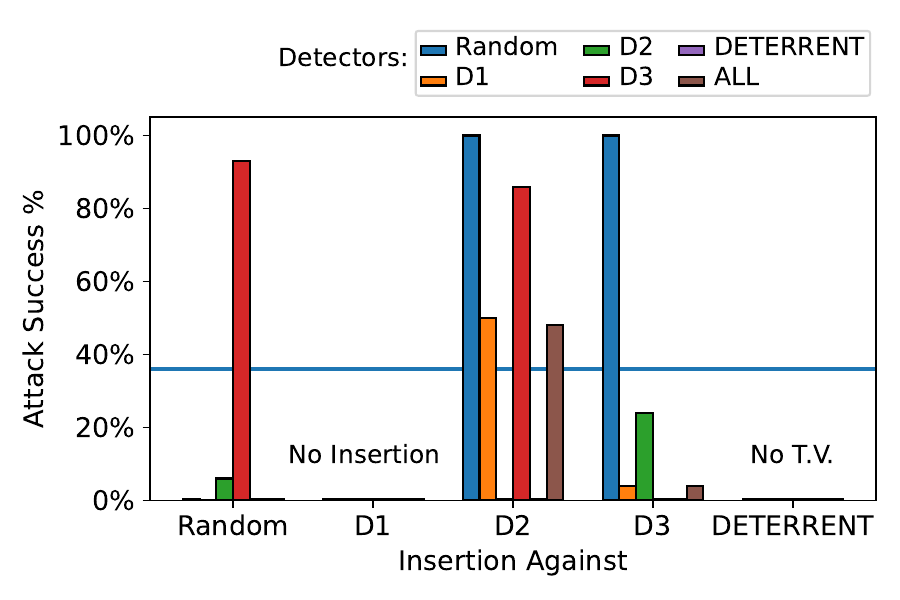}
        \caption{$c880$}
        \label{fig:c880}
    \end{subfigure}
    \hfill
    \begin{subfigure}{0.31\textwidth}
        \centering
        \includegraphics[width=\textwidth]{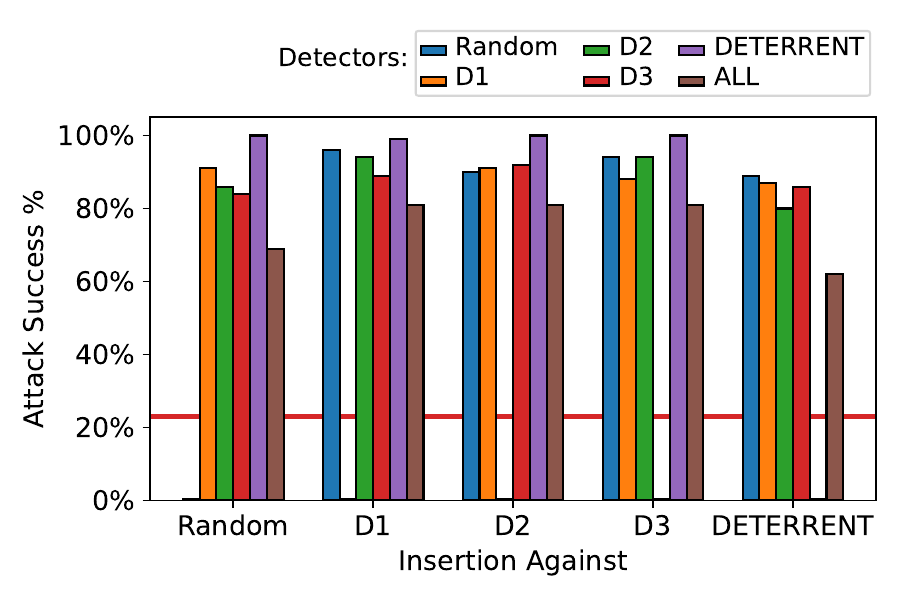}
        \caption{$c2670$}
        \label{fig:c2670}
    \end{subfigure}
    \hfill
    \begin{subfigure}{0.31\textwidth}
        \centering
        \includegraphics[width=\textwidth]{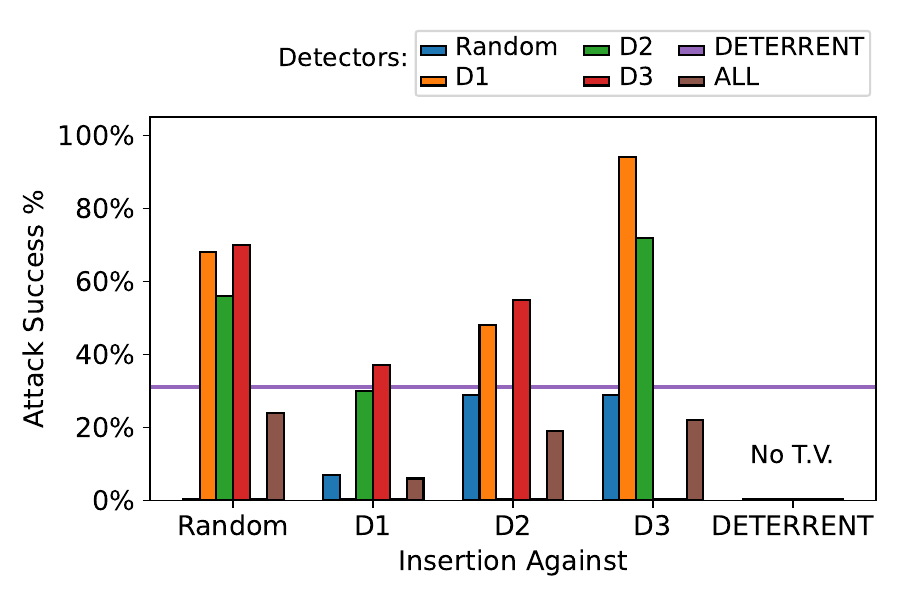}
        \caption{$c3540$}
        \label{fig:c3540}
    \end{subfigure}
    \\
    \begin{subfigure}{0.31\textwidth}
        \centering
        \includegraphics[width=\textwidth]{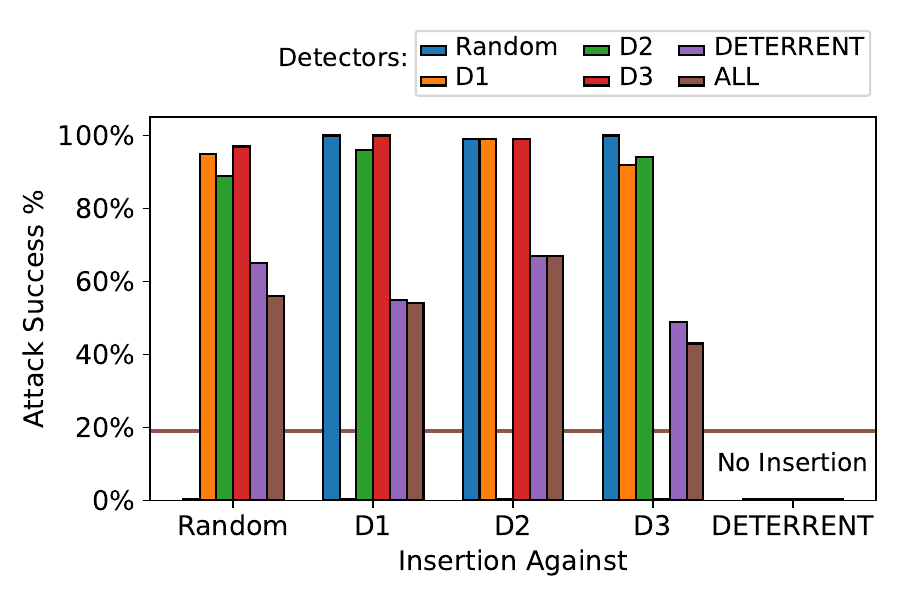}
        \caption{$c5315$}
        \label{fig:c5315}
    \end{subfigure}
    \hfill
    \begin{subfigure}{0.31\textwidth}
        \centering
        \includegraphics[width=\textwidth]{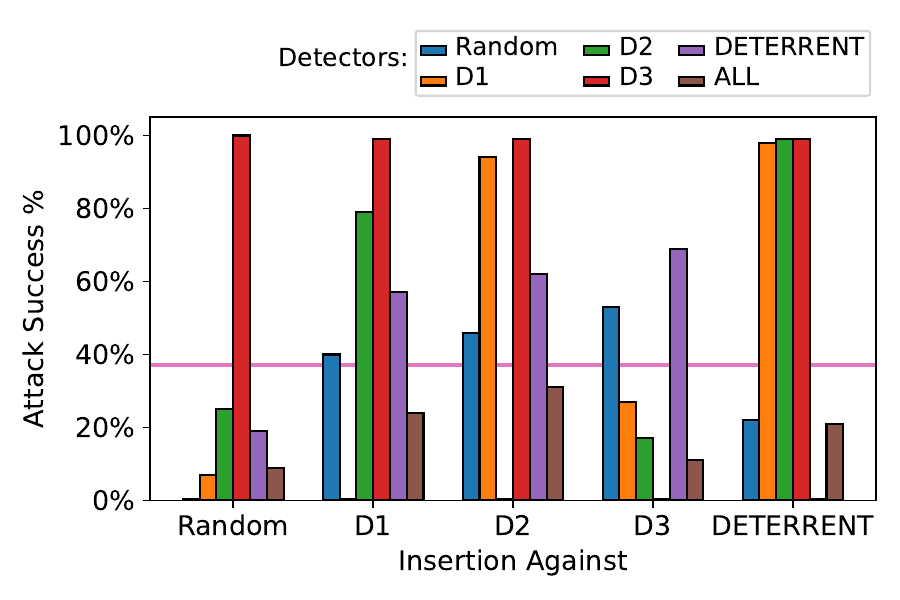}
        \caption{$c6288$}
        \label{fig:c6288}
    \end{subfigure}
    \hfill
    \begin{subfigure}{0.31\textwidth}
        \centering
        \includegraphics[width=\textwidth]{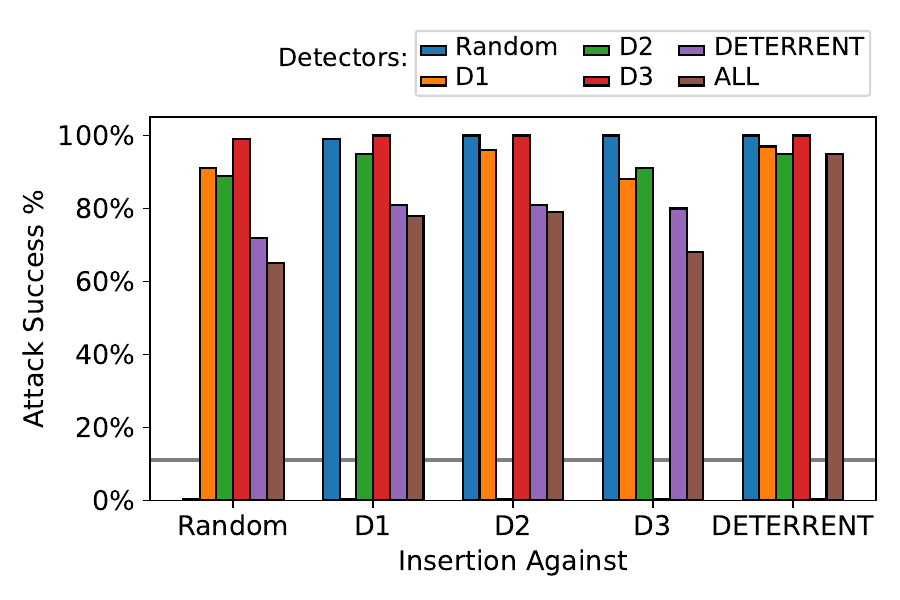}
        \caption{$c7552$}
        \label{fig:c7552}
    \end{subfigure}
    \caption{TrojanForge attack success percentages for ISCAS-85 circuits}
    \label{fig:attack}
\end{figure*}

Figure~\ref{fig:attack} shows the attack success percentages ($ASP$) of TrojanForge adversarial examples. The x-axis denotes the detectors against which the RL agent was trained, and the y-axis shows the percentage of HT instances that bypassed detection, \ie, ASP. An extra detector is denoted with \textquote{ALL} that is the combined test vectors from Random, D1, D2, D3, and DETERRENT. Also, in each of these graphs, we provide a colored horizontal line in the back that represents the $JSI$ measurement reported in Figure~\ref{fig:JSI}. We provide this visualization since it shows how effective the tool is concerning our new similarity metric.

\begin{table}[!t]
    \centering
    \caption{$c3540$ Q-coverage when $JSI$ is reduced to  to 0.17 }
    \scalebox{0.97}{
        \begin{tabular}{|c|c|c|c|c|c|}
            \hline
            \textbf{Benchmark} & \textbf{Random} & \textbf{D1} & \textbf{D2} & \textbf{D3} & \textbf{DETERRENT} \\ \hline 
            $c3540$ & 572  & 195 & 546 & 380 & N/A \\ \hline
        \end{tabular}
    }
    \label{tab:q_c3540}
\end{table}

There are two insertion cases (against D1 in $c880$ and against DETERRENT in $c5315$) where the RL agent failed (denoted by “No Insertion” in the figure). In these two cases, we hypothesize that the tool cannot insert an HT due to higher Q-coverage values and that the RL agent can not find a valid solution. In other words, TrojanForge primarily collects $0$ rewards since its insertion attempts are detected, thus failing to learn how to insert hidden HTs. There are also two other cases where test vectors from DETERRENT were not available (denoted with “No TV” in the graph). Subsequently, insertion is not feasible. Our results do not include inserting HTs into $c1355$ and $c1908$ due to the high $JSI$ and low Q-coverage (Figure~\ref{fig:JSI} and Table~\ref{tab:quintets}); for these, the tool cannot insert HTs.

Among these six circuits, insertion failed with the TrojanForge default settings in $c3540$. Accordingly, we applied the procedure in Section~\ref{subsec:JSI} to remedy the problem and reduced $JSI$ to 0.17. Table~\ref{tab:q_c3540} shows the updated Q-coverage after regular nets replaced seven out of 20 candidate nets. In this case, the random detector’s Q-coverage increases, reflecting introduced randomization stemming from adding regular nets.

As seen in Figure~\ref{fig:attack}, TrojanForge can generate successful adversarial examples. The ASP results are higher in three circuits: $c2670$, $c5315$, and $c7552$. This evidence is supported by the lower $JSI$ values for these circuits, facilitating easier HT insertions. When we insert against DETERRENT, the ASP for these circuits reaches over 60\%, 40\%, and 90\%, respectively, showing a robust tool performance. While the high Q-coverage of DETERRENT in $c5315$ leads to no insertions in this case, the ASP of \textquote{ALL} for the other four insertion strategies is well above 40\%. This underlines the importance of other less perfect detection strategies (Random, D1, D2, D3) for the success of TrojanForge. $c6288$ displays an ASP of less than 30\% in \textquote{ALL}. In this case, D1, D2, and DETERRENT have very close Q-coverage, leading to close detection rates and similar overall ASP.

\begin{figure}[t]
  \centering
  \includegraphics[scale=.35]{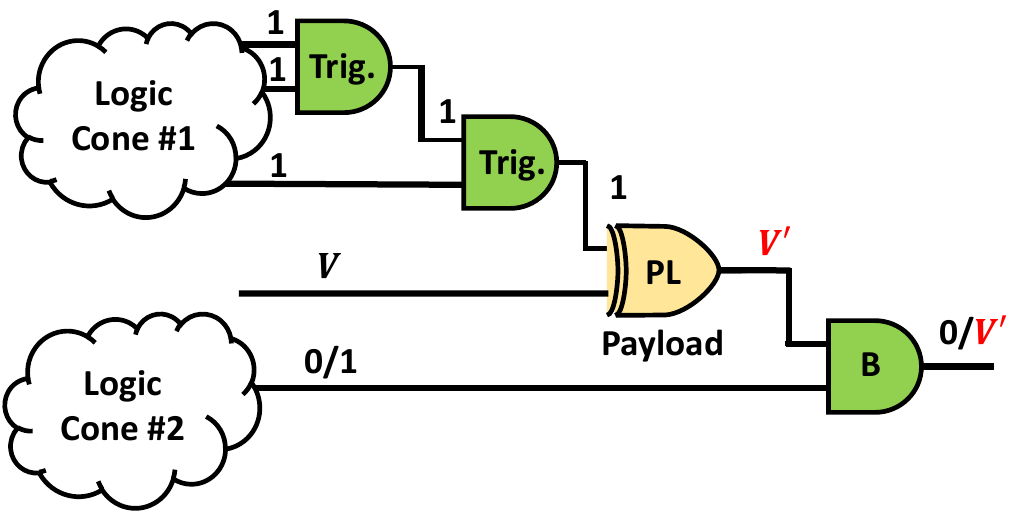}
  \caption{$JSI$ of pre-pruned candidate nets for ISCAS-85 circuits.}
  \label{fig:payload} 
\end{figure}

While the Q-coverage for DETERRENT is substantially higher than the other four detectors, there exist HT instances that remain undetected by DETERRENT when inserted against Random, D1, D2, and D3. The reason is directly related to the position of the HT payload. Figure~\ref{fig:payload} explains one such case where the trigger AND-gates are denoted with \textit{Trig.} and the payload XOR gate is denoted with \textit{PL}. Although the output of Logic Cone \#1 activates the triggers and value $V$ is ultimately inverted, $V'$ can blocked by the AND gate denoted as B if Logic Cone \#2 generates a $0$. This example demonstrates that although a detector can satisfy the triggering requirements, the choice of payload can still hide HTs. As an adversarial strategy, an attacker can connect PL’s output to a different logic cone where B is (Logic Cone \#1 $\cap$ Logic Cone \#2) and can control its value. In this strategy, the output of Logic Cone \#2 can be a net with a rare value of $1$, \ie, it is $0$ most of the time. With a specific test vector, the attacker can both activate the triggers and B to propagate the HT impact to the output, whereas when other test vectors are used, $V'$ would be masked, therefore hiding the HT. Gate B can be placed close to a circuit output for easier HT propagation.



The longest training time for TrojanForge’s flow is spent on $c6288$, and the agent requires just under six days to complete 300K timesteps. We acknowledge that this time window might be larger than a real-time attack scenario; however, we believe that our approach opens new research avenues to study GAN-like approaches and explore new attack and defense strategies in the field. We do not claim to have introduced a universal tool that is robust against all types of defense approaches. Instead, we believe our effort opens new horizons into the threat that AI-assisted adversarial tools can pose to thwart the security of electronic chips.
\section{Conclusion}
\label{sec:conclusion}
This paper presents TrojanForge, a framework that utilizes Reinforcement Learning (RL) in a GAN-like loop with Hardware Trojan (HT) detectors to insert adversarial HT instances to defeat HT detectors. Our tool applies novel pruning methods on the set of rare nets to select a smaller set of trigger candidates. An RL agent uses these candidate nets to construct HTs that can hide from a random approach and four state-of-the-art HT detectors. The RL agent reinforces actions that yield more positive rewards and generates a set of stealthy HTs in various ISCAS-85 combinational circuits. This study also investigates the impact of payload selection on the stealthiness of the HT instances and the associated security interpretations. TrojanForge can be used in the HT research space to introduce more diversity to the current HT benchmark pool and address the size and diversity issues of the current HT benchmarks. 

\section*{Acknowledgments}
This work has been partially funded by NSF grants 2219680 and 2219679.

\bibliographystyle{IEEEtran}
\bibliography{references}  
\end{document}